\documentclass[conference]{IEEEtran}

\usepackage{amsmath,amssymb}
\usepackage{graphicx}
\usepackage{booktabs}
\usepackage{array}
\usepackage{cite}
\usepackage{url}

\graphicspath{{../figures/}{figures/}}

\begin{document}

\title{Offline-Tuned Service-Aware TN--NTN Feeder Allocation for 6G Post-Disaster Service Resilience}

\author{
\IEEEauthorblockN{
Donglin Wang\IEEEauthorrefmark{2},
Anjie Qiu\IEEEauthorrefmark{2},
Qiuheng Zhou\IEEEauthorrefmark{1},
and Hans D. Schotten\IEEEauthorrefmark{2}\IEEEauthorrefmark{1}
}
\IEEEauthorblockA{\IEEEauthorrefmark{2}Rhineland-Palatinate Technical University of Kaiserslautern-Landau, Germany\\
\{dwang,qiu,schotten\}@eit.uni-kl.de}
\IEEEauthorblockA{\IEEEauthorrefmark{1}German Research Center for Artificial Intelligence (DFKI GmbH), Kaiserslautern, Germany\\
\{qiuheng.zhou,schotten\}@dfki.de}
}

\maketitle

\begin{abstract}
Terrestrial network (TN) failures during disasters can overload surviving base stations (BSs) and disrupt emergency services. A non-terrestrial network (NTN) provides fallback, but its feeder link can become a bottleneck as users migrate. This paper studies post-disaster network resilience through service-priority-aware orchestration of constrained NTN feeder capacity in a post-failure snapshot. A lightweight Cross-Entropy Method (CEM) tunes feeder, offload, and user-ranking parameters offline, not as an online artificial intelligence (AI) controller. A Python system-level simulator sweeps operating regions under abstract TN--NTN link and queueing models. Gains depend on failure severity. At a 180-Mbps feeder under total TN outage, the proposed policy increases the absolute emergency-weighted continuity metric \(C_{\mathrm{emg}}\) from 0.399 to 0.666 relative to the matched rule baseline with uniform feeder scaling (paired gain: 0.267; 95\% bootstrap CI: [0.264, 0.271]); mild damage can reduce aggregate continuity. Offline CEM feeder-weight tuning further improves severe-failure continuity over fixed service priorities, whereas tuned offload and user ranking provide no detectable incremental gain.
\end{abstract}


\section{Introduction}

Integrated terrestrial network (TN) and non-terrestrial network (NTN) systems are expected to support coverage continuity in 6G systems, especially when terrestrial infrastructure is partially unavailable. Disaster events such as floods, earthquakes, wildfires, large-scale power failures, or infrastructure attacks can disable a subset of terrestrial base stations (BSs), overload the surviving TN, and leave emergency users without reliable access. NTN fallback can restore coverage over the affected area, but the satellite feeder link and transparent-payload forwarding path impose capacity and latency constraints. Therefore, post-disaster network resilience requires not only restored coverage but also service-aware orchestration of constrained NTN feeder capacity following partial or complete terrestrial infrastructure failure.

Many TN--NTN studies evaluate average throughput, signal quality, or coverage probability. These metrics are useful but insufficient in disaster scenarios because emergency services have heterogeneous requirements. Emergency beacons require high reliability at low rate, first-responder voice requires reliable low-rate continuity, situational-awareness flows require moderate rate, and video traffic requires higher throughput but may have lower priority. A fallback policy that maximizes average rate may therefore harm emergency service continuity if it overloads the NTN feeder or migrates latency-sensitive users unnecessarily.

This paper studies service-aware TN--NTN fallback in a post-failure network snapshot. Accordingly, ``continuity'' below denotes snapshot-level application availability, not uninterrupted service over a time-evolving recovery process. The framework jointly considers heterogeneous service priorities, proactive migration, and feeder scarcity.

The primary contribution is not a new online control algorithm, but a controlled characterization of when and why service-priority-aware TN--NTN feeder allocation improves or degrades post-disaster service resilience across heterogeneous emergency services.

The main contributions are:
\begin{itemize}
    \item A system-level snapshot model combining heterogeneous emergency services, partial TN failure, constrained feeder capacity, and a class-balanced continuity metric.
    \item A demand-aware, service-priority-aware feeder allocator with iterative unused-capacity redistribution and offline CEM feeder-weight tuning; neither CEM nor weighted sharing is claimed novel.
    \item Paired Monte Carlo attribution and seed-disjoint operating-region validation exposing the crossover and lack of incremental CEM offload/ranking gain.
\end{itemize}

\section{Related Work}

\subsection{TN--NTN Resilience and NTN System Modeling}

Third Generation Partnership Project (3GPP) studies define New Radio (NR), NTN, and TN--NTN assumptions, including urban macro (UMa) propagation, transparent payloads, and procedures for long delays \cite{3gpp38901,3gpp38811,3gpp38821}. Surveys position low-, medium-, and geostationary-Earth-orbit platforms as resilient coverage enablers \cite{ntn_survey,giordani_ntn_2020}; integrated-management work emphasizes cross-domain mobility and orchestration \cite{matera_integrated_management}. Service-continuity studies advocate key performance indicators (KPIs) for access-change success, rate, latency, and application failure rather than coverage alone \cite{service_continuity}. Disaster communication studies cover aerial BSs and unmanned aerial vehicle (UAV)-assisted restoration \cite{disaster_networks,uav_disaster_survey}, but rarely combine partial TN failure, constrained NTN feeders, and heterogeneous service protection.

\subsection{Offline Optimization and Service-Aware Orchestration}

Artificial intelligence (AI)-native radio access network (RAN) and Open RAN (O-RAN) research uses data-driven control loops for traffic steering and resource allocation \cite{ai_native,oran_survey,openran_survey}. AI-RAN work studies proactive workload orchestration, while the European Space Agency's (ESA) TN--NTN white paper highlights AI-driven optimization \cite{shah_ai_ran,esa_ai_tn_ntn}. Reinforcement learning (RL), contextual bandits, and black-box optimization have also been applied to wireless resource management \cite{mnih_dqn,schulman_ppo,sutton_rl}. Here, disaster fallback is instead evaluated by emergency-weighted service success under TN damage and feeder scarcity.

\section{System Model}

\subsection{Disaster-Affected TN--NTN Network}

\begin{figure}[t]
\centering
\includegraphics[width=\columnwidth]{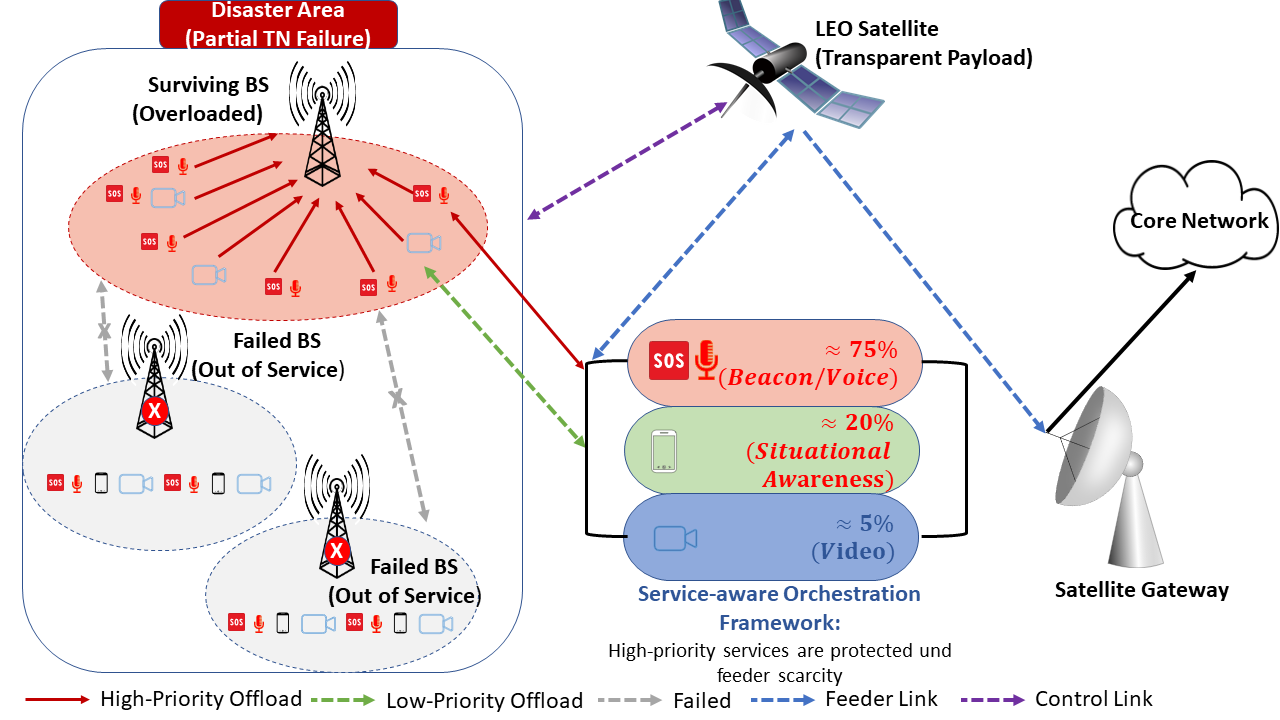}
\caption{Post-disaster TN--NTN fallback under partial TN failure and feeder scarcity. Percentages are Table~\ref{tab:services} user shares, not priority weights \(w_q\) or allocation weights \(\alpha_q\); arrows show traffic paths, not proactive-ranking order.}
\label{fig:postdisaster_architecture}
\end{figure}

We consider an integrated TN--NTN system over a disaster area, as illustrated in Fig.~\ref{fig:postdisaster_architecture}. The TN consists of \(B\) next-generation Node Bs (gNBs) serving \(K\) users. A disaster disables a subset of gNBs with probability \(p_f\), and users attached to failed gNBs are forced to re-associate or fall back to NTN. Users attached to surviving gNBs may also be proactively offloaded to NTN if the TN is overloaded or if their radio conditions are poor. The NTN layer is modeled as a LEO transparent-payload fallback link with an aggregate feeder capacity \(C_f\). When the attempted NTN fallback demand exceeds \(C_f\), feeder oversubscription reduces delivered fallback rates.

\subsection{Terrestrial and NTN Link Abstraction}

The TN side uses a 3GPP TR 38.901 UMa-inspired line-of-sight/non-line-of-sight (LOS/NLOS) path-loss abstraction at 3.5 GHz \cite{3gpp38901}. The simulator includes probabilistic LOS/NLOS state, gNB and user-equipment (UE) heights, full-frequency reuse, load-aware or signal-to-interference-plus-noise-ratio (SINR)-based association, and interference from active gNBs. The NTN access link uses a system-level link-budget abstraction with LEO slant range, antenna gains, implementation losses, and a transparent fallback path following 3GPP NTN modeling principles and our prior TN--NTN disaster simulator \cite{3gpp38811,3gpp38821,wang_disaster_simulator}. Fixed constants are inherited from these references; this paper reports the swept orchestration variables, including \(p_f\), \(C_f\), service mix, and policy settings. This abstraction is intended for rapid policy sweeps rather than packet-level protocol calibration.

\subsection{Modeling Scope}

The NTN is a transparent fallback path with an aggregate feeder constraint; orbital dynamics, Doppler tracking, beam hopping, inter-satellite routing, gateway diversity, rain fading, and protocol-level random access are omitted. Results therefore indicate orchestration trends, not absolute link performance. Each disaster is a single post-failure snapshot mapped to one fallback action; time-evolving recovery, repeated handovers, satellite passes, and Markov decision process (MDP) or online RL extensions remain future work.

\subsection{Service Model}

Each user belongs to one service class \(q\in\mathcal{Q}\). The considered service classes are emergency beacon, voice, situational awareness, and video. Each service has a minimum rate target \(R_q^{\min}\) and an emergency priority weight \(w_q\). The per-service rate-success ratio is
\begin{equation}
S_q = \frac{1}{|\mathcal{U}_q|}\sum_{u\in\mathcal{U}_q}
\mathbb{I}\left(R_u \ge R_q^{\min}\right),
\end{equation}
where \(\mathcal{U}_q\) is the user set of service \(q\), \(|\mathcal{U}_q|\) its cardinality, \(\mathbb{I}(\cdot)\) the indicator function, and \(R_u\) the delivered rate after TN/NTN selection, proactive-migration loss, and feeder scaling under oversubscription. The main continuity metric is the emergency-weighted service continuity
\begin{equation}
C_{\mathrm{emg}} =
\frac{\sum_{q\in\mathcal{Q}} w_q S_q}{\sum_{q\in\mathcal{Q}} w_q}.
\label{eq:cemg}
\end{equation}
By construction, \(C_{\mathrm{emg}}\in[0,1]\). This metric follows the service-continuity view that TN--NTN switching should be judged from the user application perspective, including delivered rate and latency impact after access change \cite{service_continuity}. Here, class-balanced means that each service contributes according to its priority rather than its number of users. Table~\ref{tab:services} defines an illustrative emergency-heavy conflict-stress configuration. The service categories and application-level continuity perspective follow \cite{service_continuity,wang_disaster_simulator}, whereas the traffic shares, delivered-rate targets, and priority weights are scenario-design parameters selected to create controlled contention under constrained NTN feeder capacity; they are not intended as standardized public-safety traffic statistics.

\begin{table}[t]
\centering
\caption{Conflict-stress service configuration.}
\label{tab:services}
\begin{tabular}{lccc}
\toprule
Service & Share & Rate target & Priority \(w_q\) \\
\midrule
Beacon & 0.30 & 0.1 Mbps & 0.40 \\
Voice & 0.45 & 0.8 Mbps & 0.45 \\
Situational awareness & 0.20 & 1.2 Mbps & 0.12 \\
Video & 0.05 & 5.0 Mbps & 0.03 \\
\bottomrule
\end{tabular}
\end{table}

\section{Offline-Tuned Service-Aware Feeder Allocation}

\subsection{Baseline Policies}

Table~\ref{tab:policy_definitions} defines the policies. Both weighted Rule variants retain Rule Uniform offload and ranking and change only \eqref{eq:feeder_allocation}. Their service-aware migration scores for beacon, voice, situational awareness, and video are \((0.1,0.3,0.6,1.0)\), respectively; higher scores increase migration preference and are combined with radio weakness. Fixed and random use the same offload budget; Static disables dynamic handover. Mandatory only retains overload-relief migration but disables discretionary weak-user migration.

\begin{table}[t]
\centering
\caption{Compared TN--NTN orchestration policies.}
\label{tab:policy_definitions}
\scriptsize
\setlength{\tabcolsep}{2.5pt}
\begin{tabular}{@{}lccc@{}}
\toprule
Policy & Offload amount & User selection & Feeder allocation \\
\midrule
CEM-Full & Tuned context & Tuned score & Service-weighted \\
Context & Learned context & Service-aware & Uniform \\
Rule Uniform & Load/weak-user rule & Service-aware & Uniform \\
Rule + Fixed Feeder & Load/weak-user rule & Service-aware & Fixed \(w_q\) \\
Rule + Tuned Feeder & Load/weak-user rule & Service-aware & CEM-tuned \\
Fixed & Fixed, \(\eta=0.5\) & Weakest first & Uniform \\
Random & Fixed, \(\eta=0.5\) & Random & Uniform \\
Mandatory only & Overload relief & Weakest excess & Uniform \\
Static & Fixed, \(\eta=0.5\) & Weakest, static & Uniform \\
\bottomrule
\end{tabular}
\end{table}

\subsection{Contextual Offload}

At each disaster snapshot, the orchestrator observes a compact state vector containing the peak surviving-gNB load ratio, the fraction of weak TN users, the estimated feeder room, the emergency-service share, and the video-service share. A learned linear score followed by a sigmoid maps this state to a proactive offload fraction \(\eta\), bounded by configured minimum and maximum values and by a feeder guard. This component controls how many candidate users are migrated. The policy is intentionally lightweight and interpretable; it is a parameterized control law trained offline, not a deep neural controller.

\subsection{Service-Aware Feeder Allocation and CEM Policy Search}

The framework first replaces uniform feeder throttling with service-weighted allocation under NTN oversubscription. Let \(D_q\) and \(C_q\) denote the aggregate fallback demand and assigned feeder capacity for service \(q\), respectively, and let \(\alpha_q\geq0\) be its allocation weight. A first-order allocation is
\begin{equation}
C_q =
\frac{\alpha_q D_q}{\sum_{j\in\mathcal{Q}}\alpha_j D_j} C_f,
\label{eq:feeder_allocation}
\end{equation}
with iterative reallocation from satisfied to unsatisfied classes. Rule + Fixed Feeder sets \(\boldsymbol{\alpha}=\mathbf{w}\) from Table~\ref{tab:services}; Rule + Tuned Feeder and CEM-Full use the normalized CEM result \((0.273,0.564,0.089,0.074)\) for beacon, voice, situational awareness, and video. This weighted demand-sharing heuristic is related to proportional-fair and max--min allocation \cite{kelly_pf,radunovic_maxmin}, but implements neither canonical optimizer nor a new fairness primitive.

Second, CEM is used to train the lightweight policy parameters. CEM has been used in wireless and LEO-satellite optimization problems because it can search high-dimensional discrete or mixed decision spaces without differentiability assumptions \cite{cem,cem_leo_reacquisition}. It learns a per-user offload score
\begin{equation}
\psi_u = \mathbf{w}_u^{T}\mathbf{x}_u,
\end{equation}
where \(\mathbf{x}_u\) contains radio weakness, SINR deficit, serving-gNB load ratio, and service-class indicators. Users with higher \(\psi_u\) are proactively migrated first.
The CEM trainer searches over contextual weights, user-score weights, and feeder-service weights. Candidate policies are evaluated by Monte Carlo simulation; elite candidates update the sampling distribution. We denote the complete CEM-tuned policy as CEM-Full. Its contextual and user-score components are retained for attribution, not assumed to be beneficial.

The CEM search uses the following scalarized offline policy-selection utility:
\begin{equation}
\begin{split}
J =\;& C_{\mathrm{emg}} + 0.015S_{\mathrm{video}}
 +0.006P_{\mathrm{video}} -0.0012L \\
& -0.10P_{\mathrm{beacon}}-0.05P_{\mathrm{voice}}-0.035\Phi ,
\end{split}
\end{equation}
where \(S_{\mathrm{video}}\) is video rate success, \(P_q\) is proactive migration ratio for service \(q\), \(L\) is average latency in ms, and \(\Phi=\max(0,D_f/C_f-1)\) is feeder oversubscription for total fallback demand \(D_f=\sum_qD_q\). This soft-constraint surrogate limits unnecessary beacon/voice migration, latency, and overload while favoring \(C_{\mathrm{emg}}\). Its auxiliary coefficients are scenario-design regularizers used only for offline selection; final comparisons use \(C_{\mathrm{emg}}\) and service KPIs. Re-scoring saved candidates under \(\pm50\%\) coefficient perturbations retained \(C_{\mathrm{emg}}\) between 0.8331 and 0.8399, a local robustness check rather than proof of optimality.
Table~\ref{tab:policy_complexity} summarizes the policy structure. Here, \(d_s\) and \(d_u\) are the state- and user-feature dimensions, \(N_{\rm eval}\) is the evaluated-candidate count, and \(T_{\rm sim}\) is the cost per simulation.

\begin{table}[t]
\centering
\caption{Policy structure and computational complexity.}
\label{tab:policy_complexity}
\begin{tabular}{lcc}
\toprule
Component & Structure & Complexity \\
\midrule
Contextual offload & Linear + sigmoid & \(O(d_s)\) \\
User scoring & Linear per user & \(O(Kd_u)\) \\
Feeder allocation & Weighted proportional & \(O(|\mathcal{Q}|)\) \\
CEM training & Black-box search & \(O(N_{\rm eval}T_{\rm sim})\) \\
\bottomrule
\end{tabular}
\end{table}

\begin{table}[t]
\centering
\caption{CEM training configuration for the conflict-stress policy.}
\label{tab:cem_training}
\begin{tabular}{lc}
\toprule
Parameter & Value \\
\midrule
Policy dimension & 18 \\
CEM iterations & 5 \\
Population per iteration & 14 candidates \\
Elite count & 4 candidates \\
Runs per candidate & 4 \\
Post-training validation & 40 runs \\
Final evaluation budget & 100 runs \\
\bottomrule
\end{tabular}
\end{table}

\begin{figure}[t]
\centering
\includegraphics[width=\columnwidth]{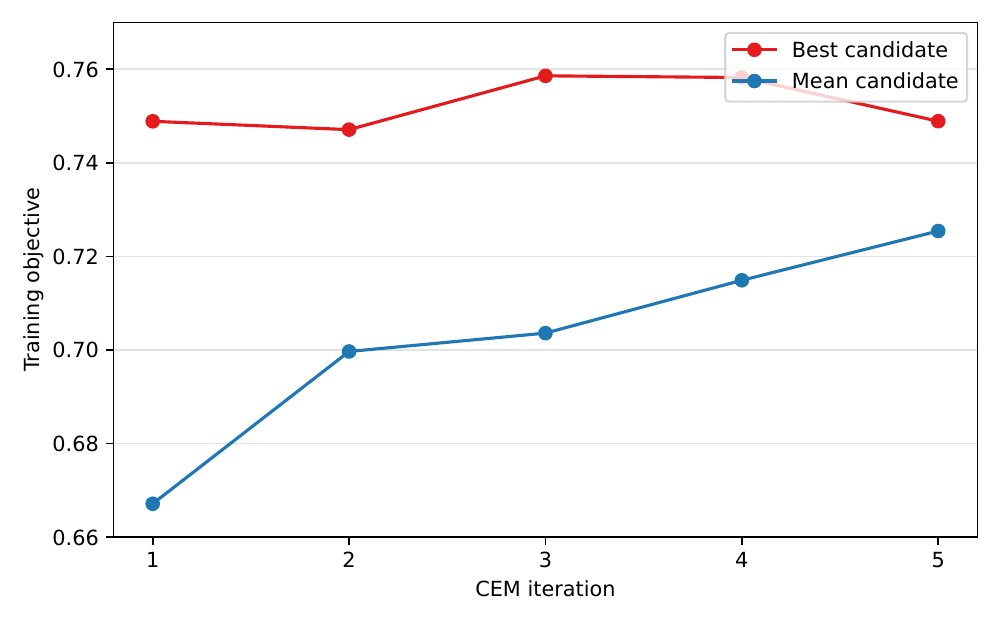}
\caption{CEM training trace. The mean candidate objective increases, while the iteration-best objective peaks in the third of five generations.}
\label{fig:cem_convergence}
\end{figure}

The CEM search uses the small reproducible budget in Table~\ref{tab:cem_training}; on an Intel Core i7-12700, an unoptimized sequential implementation takes 178 s and per-snapshot evaluation takes 0.36 ms for \(K=350\). Fig.~\ref{fig:cem_convergence} shows an increasing iteration mean; the iteration-best objective peaks in the third generation and no later candidate improves it. This is limited-budget saturation, not global convergence. Same-budget random and local Gaussian checks gave comparable validation performance, so CEM is retained as a reproducible offline tuner rather than claimed optimizer dominance. Proximal Policy Optimization (PPO) is omitted because the problem is a single-shot mapping, not a sequential MDP. Training, validation, and final-evaluation seeds are disjoint.

\section{Simulation Setup}

\subsection{System-Level Disaster Simulator}

The conflict-stress case uses \(B=10\) gNBs, \(K=350\) users, \(p_f=0.45\), and \(C_f=180\) Mbps. Rule policies use a 0.65 load threshold and \(\eta\in[0,0.6]\); CEM-Full uses \(\eta\in[0,0.65]\), and both use a 0.85 feeder guard. Main comparisons and ablations use 100 Monte Carlo runs. To avoid mixing realized severities under Bernoulli failures, a 30-run screening fixes \(N_f\in\{0,\ldots,10\}\), followed by independent 100-run paired validation at \(N_f\in\{2,4,6,8,10\}\) with disjoint seeds. We track \(C_{\mathrm{emg}}\), service success, latency, and feeder indicators. Common random numbers preserve user, service, failed-gNB, and channel realizations across policies. Paired differences use 95\% bootstrap intervals, two-sided Wilcoxon signed-rank tests with tie correction, and Holm correction.

\section{Results}

\subsection{NTN Fallback Benefit}

\begin{figure}[t]
\centering
\includegraphics[width=\columnwidth]{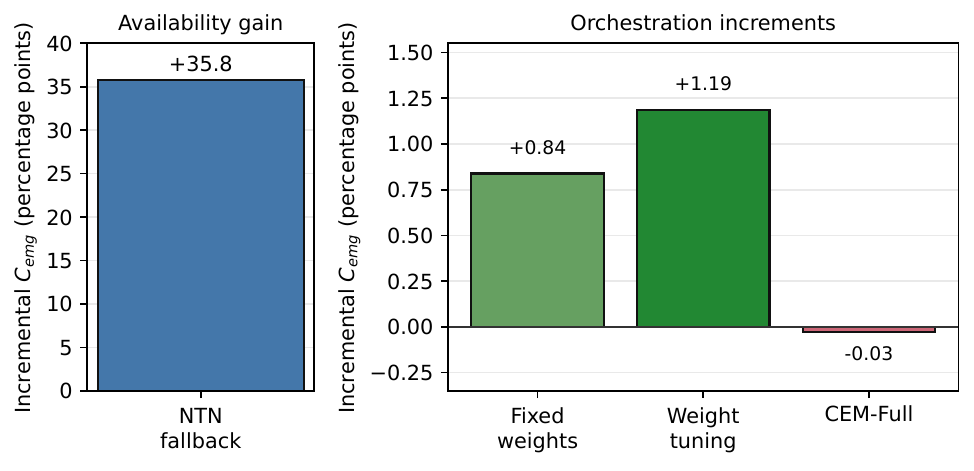}
\caption{Incremental \(C_{\mathrm{emg}}\) attribution in percentage points under conflict stress; separate panel scales expose the smaller orchestration effects.}
\label{fig:component_attr}
\end{figure}

Before comparing orchestrators, Fig.~\ref{fig:component_attr} isolates the basic value of NTN access. NTN fallback raises \(C_{\mathrm{emg}}\) by 35.8 percentage points over TN-only recovery, establishing the fallback layer as the dominant availability contribution. It does not remove feeder oversubscription; the right panel instead quantifies how allocation mitigates its service-level impact.

\subsection{Service-Priority-Aware Feeder Allocation}

With the offload rule and service-aware ranking held fixed, the right panel of Fig.~\ref{fig:component_attr} shows that Rule + Fixed Feeder adds 0.84 points over Rule Uniform, while offline CEM feeder-weight tuning in Rule + Tuned Feeder adds another 1.19 points. Thus, service-priority-aware allocation and its weight tuning account for the measurable orchestration gain.

\begin{figure}[t]
\centering
\makebox[\columnwidth][c]{\includegraphics[width=1.10\columnwidth]{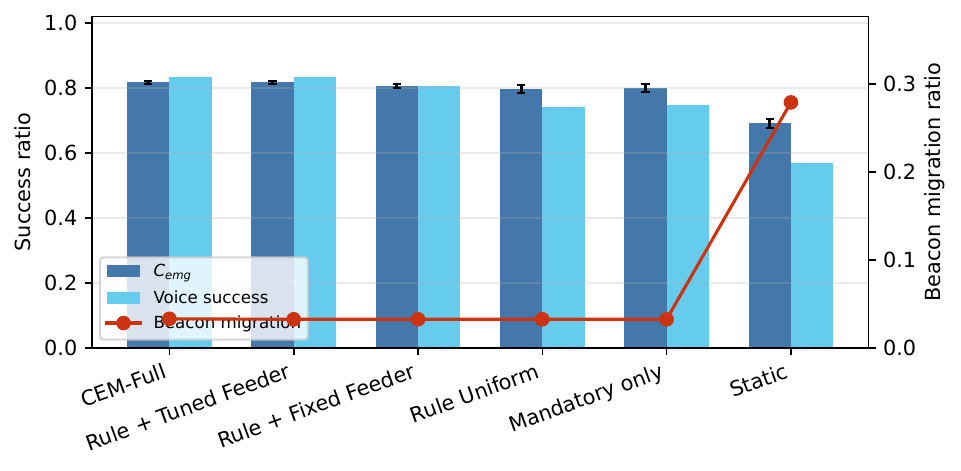}}
\caption{Conflict-stress comparison with 350 users, \(p_f=0.45\), and \(C_f=180\) Mbps. Error bars are 95\% normal-approximation intervals for mean \(C_{\mathrm{emg}}\); the line is the beacon fraction migrated from surviving TN, including mandatory overload relief.}
\label{fig:conflict}
\end{figure}

Fig.~\ref{fig:conflict} reports absolute performance over the same 100 snapshots. Rule + Tuned Feeder, Rule + Fixed Feeder, Rule Uniform, Mandatory only, and Static obtain \(C_{\mathrm{emg}}\) values of 0.8171, 0.8053, 0.7969, 0.8001, and 0.6908; CEM-Full obtains 0.8168 and is analyzed next. Voice success is 0.8326, 0.8060, and 0.7428 for the three matched Rule policies. Static aggressive migration is harmful under feeder scarcity.

\subsection{CEM Attribution and Ablation}

Replacing Rule + Tuned Feeder with CEM-Full changes \(C_{\mathrm{emg}}\) by only \(-0.03\) points in Fig.~\ref{fig:component_attr}; their absolute values are 0.8168 and 0.8171. The paired difference is \(-0.0003\), with a bootstrap 95\% interval of [\(-0.0022\), 0.0016], so tuned offload and user ranking provide no detectable incremental gain at this operating point.

The 100-run ablations are CEM-Full (context C, user score U, feeder allocation F), User Score (C+U), Feeder (C, fixed ranking, F), Context (C, fixed ranking), and Rule Uniform (rule offload); variants without F use uniform scaling. Their mean \(C_{\mathrm{emg}}\) values \(\pm\) 95\% normal-approximation half-width are \(0.8168\pm0.0044\), \(0.7957\pm0.0126\), \(0.8171\pm0.0045\), \(0.7963\pm0.0125\), and \(0.7969\pm0.0126\). Feeder allocation is dominant, while CEM-Full and Feeder are practically indistinguishable; hence no consistent user-scoring gain is established.

\subsection{Failure-Severity Region}

\begin{figure}[t]
\centering
\makebox[\columnwidth][c]{\includegraphics[width=1.08\columnwidth]{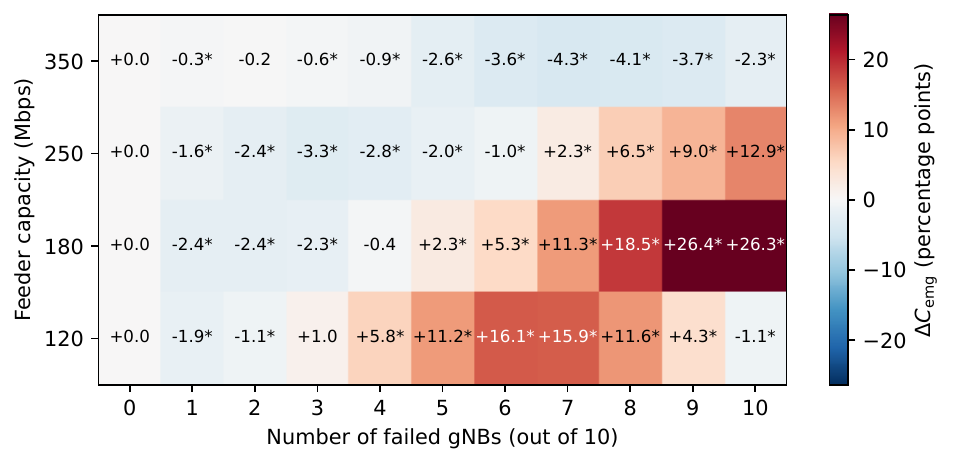}}
\caption{Failure-severity and feeder-capacity screening. Cells show the paired mean change in \(C_{\mathrm{emg}}\), in percentage points, for Rule + Tuned Feeder minus Rule Uniform over 30 common-random-number snapshots. An asterisk denotes a bootstrap 95\% interval excluding zero and Holm-adjusted Wilcoxon \(p<0.05\).}
\label{fig:failure_capacity_grid}
\end{figure}

Fig.~\ref{fig:failure_capacity_grid} screens \(N_f\in\{0,\ldots,10\}\) and \(C_f\in\{120,180,250,350\}\) Mbps. The screening suggests that the sign and significance of the Rule + Tuned Feeder gain depend jointly on damage and feeder capacity; no single failure-count threshold appears universal. Because each cell uses only 30 runs, the grid is treated as operating-region screening, while Tables~\ref{tab:operating_paired}--\ref{tab:service_tradeoff} provide independent 100-run confirmation at selected points.

Table~\ref{tab:operating_paired} gives independent 100-run checks at \(C_f=180\) Mbps. Relative to Rule Uniform, Rule + Tuned Feeder changes \(C_{\mathrm{emg}}\) by \(-2.49\), \(-0.58\), +5.84, +16.57, and +26.73 percentage points at \(2/10\), \(4/10\), \(6/10\), \(8/10\), and \(10/10\) failed gNBs, respectively, revealing a clear operating-region crossover. Rule + Fixed Feeder gains 3.98 points at \(6/10\) and remains below tuned allocation at the three severe points. CEM-Full minus Rule + Tuned Feeder is 0.09, 0.02, 0.07, \(-0.06\), and 0.00 points; every bootstrap interval includes zero. Because all users require NTN at \(10/10\), equality there isolates the benefit to feeder allocation rather than offload or ranking.

\begin{table}[t]
\centering
\caption{Exact-failure validation of absolute \(C_{\mathrm{emg}}\) at \(C_f=180\) Mbps.}
\label{tab:operating_paired}
\scriptsize
\setlength{\tabcolsep}{3pt}
\begin{tabular}{ccccc}
\toprule
\(N_f/10\) & Rule Uniform & Fixed Feeder & Tuned Feeder & CEM-Full \\
\midrule
2/10 & 0.8438 & 0.8193 & 0.8189 & 0.8198 \\
4/10 & 0.8296 & 0.8162 & 0.8238 & 0.8240 \\
6/10 & 0.7516 & 0.7914 & 0.8100 & 0.8107 \\
8/10 & 0.5936 & 0.7215 & 0.7592 & 0.7586 \\
10/10 & 0.3987 & 0.5783 & 0.6660 & 0.6660 \\
\bottomrule
\end{tabular}
\end{table}

Table~\ref{tab:service_tradeoff} explains the sign reversal over paired snapshots. Win rate \(W_{\mathrm{tuned}}=100N^{-1}\sum_r\mathbb{I}(C_{\mathrm{emg},r}^{\mathrm{tuned}}>C_{\mathrm{emg},r}^{\mathrm{uniform}})\) is the percentage won by tuned allocation; no ties occur. At \(2/10\), tuned changes voice and situational awareness by +1.58 and \(-26.65\) points and never wins; at \(6/10\), the changes are +13.79 and \(-2.93\) points and every snapshot wins. At \(8/10\) and \(10/10\), voice changes by +37.53 and +61.58 points, beacon by \(-0.80\) and \(-2.45\), the other services are unchanged, and every snapshot wins. Thus, severe-region gains come primarily from voice protection, while \(\rho_f\in[2.19,2.30]\) shows that feeder load alone cannot determine activation.

\begin{table}[t]
\centering
\caption{Rule + Tuned Feeder minus Rule Uniform: paired service trade-off at \(C_f=180\) Mbps. Changes are percentage points.}
\label{tab:service_tradeoff}
\scriptsize
\setlength{\tabcolsep}{1.5pt}
\begin{tabular}{ccccccc}
\toprule
\(N_f/10\) & \(\rho_f\) & \(\Delta C_{\mathrm{emg}}\) & \(\Delta S_{\mathrm{bea.}}\) & \(\Delta S_{\mathrm{voi.}}\) & \(\Delta S_{\mathrm{sit.}}\) & Win \\
\midrule
2/10 & 2.19 & \(-2.49\) & \(0.00\) & \(+1.58\) & \(-26.65\) & 0\% \\
4/10 & 2.23 & \(-0.58\) & \(0.00\) & \(+5.42\) & \(-25.12\) & 29\% \\
6/10 & 2.25 & \(+5.84\) & \(-0.03\) & \(+13.79\) & \(-2.93\) & 100\% \\
8/10 & 2.25 & \(+16.57\) & \(-0.80\) & \(+37.53\) & \(0.00\) & 100\% \\
10/10 & 2.30 & \(+26.73\) & \(-2.45\) & \(+61.58\) & \(0.00\) & 100\% \\
\bottomrule
\end{tabular}
\end{table}

\section{Discussion and Limitations}

The results support a narrower claim than an adaptive-AI controller: demand-aware feeder allocation protects priority services under scarcity, with CEM used for offline feeder-weight tuning. Tuned offload and user ranking add no benefit. Service parameters and utility coefficients are illustrative rather than standardized; their local perturbation test does not establish generalization across disaster types.

\subsection{Design Insight}

Weighted allocation should not be permanently enabled. Table~\ref{tab:service_tradeoff} shows that similar \(\rho_f=D_f/C_f\) load ratios yield opposite effects, so a threshold based only on failure count or feeder load cannot identify the crossover. Safe switching requires class-specific unmet demand and marginal success per allocated Mbps; designing that controller remains future work.

The trade-off is consistent with capacity redistribution: Rule + Tuned Feeder favors voice users near their 0.8-Mbps target while some situational-awareness users fall below 1.2 Mbps. Severe damage makes the voice recovery outweigh this displacement; mild damage does not.

Table~\ref{tab:service_tradeoff} also exposes a scalar-objective risk: with weight 0.12, a 26.65-point situational-awareness loss appears as only a 2.49-point \(C_{\mathrm{emg}}\) decrease. Safe deployment should impose \(S_q(\pi)\geq S_q^{\mathrm{floor}}\) for every safety-relevant class while maximizing \(C_{\mathrm{emg}}\) over the three Rule feeder variants. This constraint is motivated, not evaluated.

\subsection{External-Validation Scope}

All feeder results come from the authors' system-level simulator; no packet-level or third-party validation covers the sign reversal. Detailed NR scheduling, hybrid automatic repeat request (HARQ), radio link control (RLC), radio resource control (RRC) signaling, NTN random access, and orbital dynamics are omitted. The results are controlled comparative evidence, not protocol-level prediction.

CEM is an offline search for a static decision, not a temporal controller. Missing greedy, mathematical-programming, and contextual-bandit baselines limit algorithmic claims; strong baselines, multiple service mixes, and time-evolving recovery are required next steps.

\section{Conclusion}

This paper studied post-disaster service resilience in TN--NTN snapshots. Under conflict stress, NTN fallback adds 35.8 percentage points over TN-only recovery; service-priority-aware feeder allocation then has operating-region-dependent effects. At 180 Mbps, Rule + Tuned Feeder changes emergency-weighted continuity by \(-2.49\), \(-0.58\), +5.84, +16.57, and +26.73 points from \(2/10\) through \(10/10\) failed gNBs; voice gains reach 13.79, 37.53, and 61.58 points at the three severe points. CEM-Full matches Rule + Tuned Feeder, showing no incremental gain from CEM-tuned offload or user ranking. Weighting can harm situational-awareness demand, and total feeder load does not identify this reversal. Conclusions remain limited to the abstract model and illustrative service mix. Future work will add strong baselines, multi-scenario and temporal evaluation, safe switching, and independent packet-level TN--NTN validation.

\section*{Acknowledgment}

This work was supported by the German Federal Ministry of Research, Technology and Space (BMFTR) through Open6GHub+ (16KIS2406), X-COM (16KISS007K), and 6G-Coverage (16KIS2425).

\IEEEtriggeratref{6}

\end{document}